\documentclass[prl,letterpaper,twocolumn,showpacs,showkeys,lengthcheck,
               floatfix,nofootinbib,preprintnumbers,superscriptaddress]{revtex4-1}
\usepackage{amsmath,amssymb,amsfonts,graphicx}
\usepackage{bm}
\usepackage{hyperref}
\usepackage{slashed}
\usepackage[caption=false]{subfig}
\usepackage[svgnames]{xcolor}
\usepackage[english]{babel}
\usepackage{blindtext}
\usepackage{microtype}
\usepackage{tikz}
\RequirePackage{slashed}

\def\g{\gamma}

\def\no{\nonumber}

\def\be{\begin{equation}}
\def\ee{\end{equation}}

\begin{document}


\title{Electromagnetic Contribution to the Proton-Neutron Mass Splitting}

\author{A. W. Thomas}
\affiliation{ARC Centre of Excellence for Particle Physics at the Terascale and CSSM, School of Chemistry and Physics, University of Adelaide 5005, Australia.}
\author{X.~G.~Wang}
\affiliation{ARC Centre of Excellence for Particle Physics at the Terascale and CSSM, School of Chemistry and Physics, University of Adelaide 5005, Australia.}
\author{R. D. Young}
\affiliation{ARC Centre of Excellence for Particle Physics at the Terascale and CSSM, School of Chemistry and Physics, University of Adelaide 5005, Australia.}

\begin{abstract}
We study the electromagnetic contribution to the proton-neutron mass splitting 
by combining lattice simulations and the modified Cottingham sum rule of 
Walker-Loud, Carlson and Miller. This analysis yields an estimate of the 
isovector nucleon magnetic polarizability as a function of pion mass. 
The physical value, obtained by chiral extrapolation to the physical pion mass, 
is $\beta_{p-n}=(-1.12 \pm 0.40)\times 10^{-4}\ \mathrm{fm}^3$, 
which is in agreement with the empirical result, albeit with a somewhat 
smaller error. As a result, 
we find $\delta M^{\gamma}_{p-n}=1.04 \pm 0.11\ \mathrm{MeV}$, 
which represents a significant improvement in precision.
\end{abstract}


\maketitle

%
The physical proton-neutron mass splitting has been measured extremely 
precisely~\cite{Mohr:2012,PDG:2012},
\begin{equation}
M_n - M_p =1.2933322(4)\ \mathrm{MeV}\ .
\end{equation}
Its separation into contributions from electromagnetic effects and 
the $u-d$ quark mass difference is of enormous 
interest~\cite{Zee:1972,Leutwyler:1982,Miller:1990,Shanahan:2013}. 
Not only are the $u-d$ masses critical parameters in the study of explicit 
chiral symmetry breaking in QCD but their precise values are vital to the 
discussion of mass generation (within the framework of grand unification),
as well as the mechanism of $CP$ violation. Clearly, if one of the two components 
of $M_n-M_p$ can be determined accurately, the other may be inferred from the total.

Recently, Walker-Loud {\it et al.}(WLCM)~\cite{Walker-Loud:2012} showed
how to use the formal operator product expansion (OPE) 
analysis of Collins~\cite{Collins:1979}
to overcome an ambiguity in the original approach of Cottingham~\cite{Cottingham:1963}.
The WLCM analysis led to a significantly larger numerical value for 
the electromagnetic contribution to the p-n mass difference
but with a rather large uncertainty,
\begin{equation}
\delta M^{\gamma}|_{p-n}=1.30(03)(47)\ \mathrm{MeV}\ .
\end{equation}
This may be compared with the value of Gasser and Leutwyler, based on the 
standard Cottingham sum-rule, of $0.76(30)\ \mathrm{MeV}$~\cite{Leutwyler:1982}.

An alternative approach to this problem involves the direct calculation of 
the electromagnetic contribution to the mass shift using lattice 
QCD~\cite{Duncan:1996,Blum:2010}. Most recently, the 
BMW Collaboration~\cite{BMW:2013} reported a value of $1.59(30)(35)\ \mathrm{MeV}$.

In the WLCM formalism, the total electromagnetic contribution to 
the p-n mass shift, denoted by $\delta M^{\gamma}$, 
is written as the sum of five terms,
\begin{equation}\label{eq:fullsplitting}
\delta M^{\g} = \delta M^{el} + \delta M^{inel} + \delta M^{el}_{sub} + \delta M^{inel}_{sub} + \delta\tilde{M}^{ct}\ .
\end{equation}
The terms $\delta M^{el}$, $\delta M^{inel}$ and $\delta M^{el}_{sub}$ are 
uncontroversial and can be evaluated very accurately. Using the Kelly 
parametrization of the nucleon electromagnetic form factors~\cite{Kelly:2004} 
and modern knowledge of the structure functions~\cite{Christy:2008,Christy:2010}, 
one finds~\cite{Walker-Loud:2012}
\begin{equation}\label{eq:3parts}
(\delta M^{el}+\delta M^{inel}+\delta M^{el}_{sub}+\delta\tilde{M}^{ct})|_{p-n}=
0.83 \pm 0.04\ \mathrm{MeV}\ ,
\end{equation}
where we have combined the uncertainties in quadrature. 
(Following WLCM, the counter term, $\delta\tilde{M}^{ct}$, which is related to 
the $\pi$-N sigma commutator is set to zero with an uncertainty 
of $\pm 0.02\ \mathrm{MeV}$.) On the other hand, the inelastic subtraction term, 
which as a matter of principle should be considered together with 
the very small counter term
introduced by Collins, has been reported as
\begin{equation}
\delta M^{inel}_{sub}|_{p-n}=0.47 \pm 0.47\ \mathrm{MeV}\ .
\end{equation}
This large uncertainty, which in turn dominates the overall uncertainty on 
$\delta M^{\g}$, results from the uncertainty associated with the isovector nucleon 
magnetic polarizability, which was taken to be
\begin{equation}
\beta_{p-n} = (-1\pm 1)\times 10^{-4}\ \mathrm{fm}^3\ .
\end{equation}

In this Letter, we use data from the RBC Collaboration~\cite{Blum:2010} for 
the electromagnetic mass shift as a function of quark mass to provide an 
improved constraint on the inelastic subtraction term of WLCM. 
In this way, the contribution from the term involving the 
isovector nucleon magnetic polarizability can be extracted as a function of 
pion mass. The result is a considerable improvement in the precision of 
the overall electromagnetic contribution to the mass splitting.

\begin{figure}[t]
\centering\includegraphics[width=\columnwidth,clip=true,angle=0]{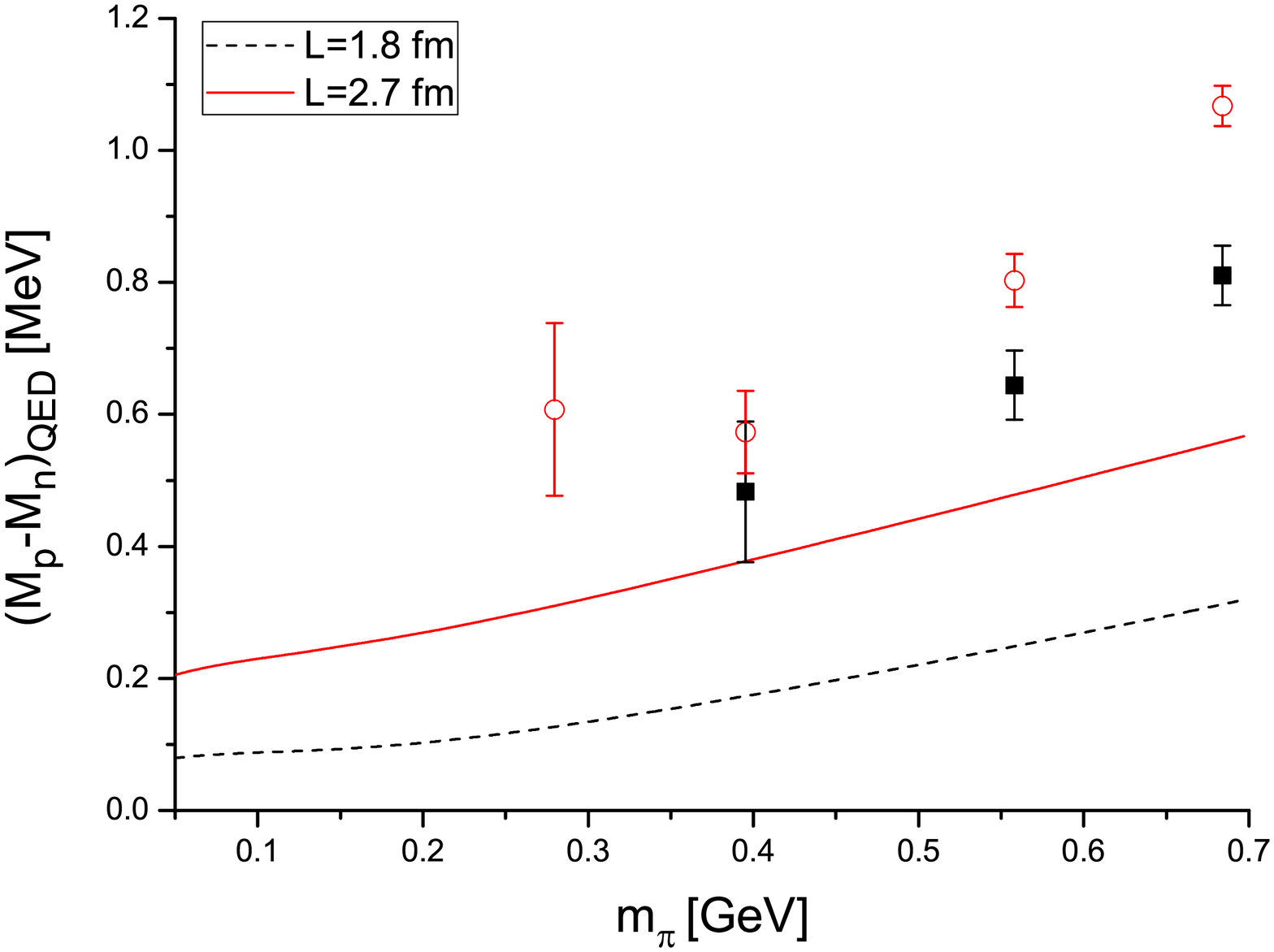}
\caption{Total elastic contribution to nucleon mass splitting at finite volume. 
Lattice data are taken from Ref.~\cite{Blum:2010}, $16^3$ (squares) 
and $24^3$ (circles) lattice sizes.}
\label{fig:chiral-extrapolation}
\end{figure}

To begin, we consider the finite volume lattice QCD calculation of the 
RBC Collaboration~\cite{Blum:2010}. This group has reported the 
electromagnetic mass difference as a function of quark mass for two 
lattice volumes, $16^3$ ($L=1.8\ \mathrm{fm}$) 
and $24^3$ ($L=2.7\ \mathrm{fm}$) with lattice 
cutoff $a^{-1}\approx 1.78\ \mathrm{GeV}$.
In Fig.~\ref{fig:chiral-extrapolation}, we compare their results for the 
electromagnetic p-n mass splitting with the finite volume 
versions~\cite{Davoudi:2014qua} 
\begin{eqnarray}\label{eq:finite-volume}
\delta M^{el}(L) &=& \frac{2\pi\alpha}{L^3}\sum_{\overrightarrow{q}\neq0}\frac{1}{Q^2}
\left\{ \frac{ 3\sqrt{\tau_{el}} G^2_M}{2(1+\tau_{el})} + \frac{[G^2_E-2\tau_{el}G^2_M]}{1+\tau_{el}} \right.\nonumber\\
&&\left.\times \left[ 1+\tau_{el})^{3/2} - \tau^{3/2}_{el} - \frac{3}{2}\sqrt{\tau_{el}} \right] \right\}\ ,\nonumber\\
\delta M^{el}_{sub}(L) &=& -\frac{3\alpha\pi}{4M}\frac{1}{L^3}\sum_{\overrightarrow{q}\neq0}\frac{1}{Q}[2G^2_M - 2F^2_1]
\end{eqnarray}
of the infinite volume expressions for the elastic 
contributions, $\delta M^{el} + \delta M^{el}_{sub}$:
\begin{eqnarray}
\delta M^{el} &=& \frac{\alpha}{\pi}\int^{\Lambda_0}_{0}dQ
\left\{ \frac{ 3\sqrt{\tau_{el}} G^2_M}{2(1+\tau_{el})} + \frac{[G^2_E-2\tau_{el}G^2_M]}{1+\tau_{el}} \right.\nonumber\\
&&\left.\times \left[ 1+\tau_{el})^{3/2} - \tau^{3/2}_{el} - \frac{3}{2}\sqrt{\tau_{el}} \right] \right\}\ ,\nonumber\\
\delta M^{el}_{sub} &=& -\frac{3\alpha}{16\pi M}\int_0^{\Lambda^2_0}dQ^2[2G^2_M - 2F^2_1]\ ,
\end{eqnarray}
with $Q=|\overrightarrow{q}|$ (in the heavy baryon limit) 
and $\tau_{el}=Q^2/4M^2$.

%
In order to evaluate Eqs.~(\ref{eq:finite-volume}) as a function of quark, 
or equivalently pion mass,
we use a parametrization of lattice data for the nucleon isovector 
and isoscalar form factors introduced in Ref.~\cite{Ashley:2004}. That is, we use
\begin{eqnarray}
G^{v,s}_{M}=\frac{\mu_{v,s}(m_{\pi})}{ (1 + Q^2/(\Lambda^{v,s}_M)^2 )^2}\ ,\nonumber\\
G^{v,s}_{E}=\frac{1}{ (1 + Q^2/(\Lambda^{v,s}_E)^2 )^2}\ .
\label{eq:GMGE}
\end{eqnarray}
Following~\cite{Thomas:1999,Thomas:2000}, one can use a Pad\'e approximant 
to quite accurately parametrize the magnetic moments
\begin{equation}
\mu_i(m_{\pi})=\frac{\mu_0}{1-\frac{\chi_i}{\mu_0}m_{\pi}+c m_{\pi}^2}\ .
\end{equation}
The chiral coefficients for the isovector and isoscalar moments have been fixed  
at the values required by chiral perturbation 
theory~\cite{Thomas:2000}, 
$\chi_v=-(F+D)^2 m_N/4\pi f_{\pi}^2 = -8.82\ \mathrm{GeV}^{-1}$ and $\chi_s=0$. 
The parameters $(\mu_0, c)$, in units of $(\mu_N,\mathrm{GeV}^{-2})$, 
are determined by fitting lattice data~\cite{QCDSF:2005}, 
$\mu_0=5.71$, $c=-0.21$ for the isovector moments and $\mu_0=0.86$, $c=0.55$ for 
the isoscalar moments, respectively.

The dipole masses of the isovector magnetic and electric form factors are 
parametrized as
\begin{eqnarray}
(\Lambda^v_M)^2 &=& \frac{12 ( 1 + A_1 m_{\pi}^2 )}{ A_0 + \frac{\chi_1}{m_{\pi}} 
\frac{2}{\pi} \arctan(\mu/m_{\pi}) + \frac{\chi_2}{2} 
\ln(\frac{m_{\pi}^2}{m_{\pi}^2+\mu^2}) }\ ,\nonumber\\
(\Lambda^v_E)^2 &=& \frac{12 ( 1 + B_1 m_{\pi}^2 )}{ B_0 + \frac{\chi_2}{2} 
\ln(\frac{m_{\pi}^2}{m_{\pi}^2+\mu^2}) }\ ,
\end{eqnarray}
where, once again, the leading non-analytic behavior of the charge and magnetic 
radii are chosen to agree with chiral perturbation theory:
\begin{equation}
\chi_1 = \frac{g_A^2 m_N}{8\pi f_{\pi}^2 \kappa_v}\ ,\ \ \ \ \ \chi_2 = -\frac{5g_A^2 + 1}{8\pi^2 f_{\pi}^2}\ ,
\end{equation}
with $g_A=1.27$ the axial coupling constant and $f_{\pi}=93\ \mathrm{MeV}$ 
the pion decay constant. $m_N=940\ \mathrm{MeV}$ is the nucleon mass 
and $\kappa_v=4.2$ is the isovector anomalous magnetic moment of 
the nucleon (in the chiral limit).

The isoscalar dipole masses are observed to be roughly linear in $m_{\pi}^2$,
\begin{equation}
(\Lambda^s_{E,M})^2=a_{E,M} + b_{E,M}m_{\pi}^2\ .
\end{equation}
The parameters are determined by fitting lattice data from 
the QCDSF Collaboration~\cite{QCDSF:2005},
\begin{eqnarray}
A_0 &=& 8.65\ ,\ \ \ \ A_1=0.28\ ,\nonumber\\
B_0 &=& 11.71\ ,\ \ \ B_1=0.72\ ,
\end{eqnarray}
in units of $\mathrm{GeV}^{-2}$ and
\begin{eqnarray}
a_E &=& 1.09\ \mathrm{GeV}^2,\ \ \ \ b_E=0.85\ ,\nonumber\\
a_M &=& 1.09\ \mathrm{GeV}^2,\ \ \ \ b_M=0.68\ ,
\end{eqnarray}
with $\mu=0.14\ \mathrm{GeV}$. The electromagnetic form factors of 
the proton and neutron can be reconstructedthrough  
\begin{equation}
G^p=\frac{1}{2}(G^s + G^v)\ ,\ \ G^n=\frac{1}{2}(G^s - G^v)\ .
\end{equation}

Figure~\ref{fig:chiral-extrapolation} shows the total elastic contribution 
to the proton-neutron mass difference, computed at two lattice volumes.
While the general behavior of the calculation is in agreement with 
the lattice simulations, there is a clear discrepancy. 
This discrepancy is identified with the finite volume version of the 
inelastic contribution, $\delta M^{inel}+\delta M^{inel}_{sub}$ 
in Eq.~(\ref{eq:fullsplitting}). Since $\delta M^{inel}$ is so small, 
we expect that any finite volume corrections to it will be well within 
the uncertainty quoted by WLCM and therefore we simply include the 
physical value ($\delta M^{inel}=0.057\ \mathrm{MeV}$) in our calculation.

\begin{table*}[t]
\begin{center}
\caption{\label{tab:beta-mpi} The magnetic polarizability $\beta_{p-n}$ as a function of $m_{\pi}$, in units of $10^{-4}\ \mathrm{fm}^3$. $\delta M^{inel}_{sub}$ is given by Eq.~(\ref{eq:inelastic-subtraction-term}).}
\renewcommand{\arraystretch}{1.3}
\begin{tabular}{c|c|cccc}\hline
  \      &$m_{\pi} [\mathrm{GeV}]$    &       $0.279$        &       $0.394$         &         $0.558$          &       $0.683$        \\ \hline
 $n=3$   &        $16^3$              &          \           &   $-0.246 \pm 0.103$  &  $-0.258 \pm 0.040$      &  $-0.294 \pm 0.030$  \\
  \      &        $24^3$              &  $-0.316 \pm 0.171$  &   $-0.134 \pm 0.060$  &  $-0.202 \pm 0.030$      &  $-0.298 \pm 0.020$ \\ \hline
 $n=4$   &        $16^3$              &          \           &   $-0.733 \pm 0.307$  &  $-0.756 \pm 0.118$      &  $-0.855 \pm 0.087$  \\
   \     &        $24^3$              &  $-0.917 \pm 0.498$  &   $-0.385 \pm 0.172$  &  $-0.578 \pm 0.087$      &  $-0.847 \pm 0.057$  \\ \hline
\end{tabular}
\end{center}
\end{table*}
Turning to the inelastic subtraction term, we note that the dipole form factor 
multiplying $\beta_{p-n}Q^2$, which was used by WLCM, leads to a very 
large $\log(Q^2_0)$ term. The magnitude is inconsistent with the $\log(Q^2_0)$ 
behavior of $\delta\tilde{M}^{ct}$, which is the only term that will contribute 
to the asymptotic scaling. In order to avoid this problem, we choose to 
use a form factor with either cubic or quartic behavior. Thus the finite 
volume corrections missing in Fig.~\ref{fig:chiral-extrapolation} are taken as:
\begin{equation}\label{eq:inelastic-subtraction-term}
\delta M^{inel}_{sub}(L) = -\frac{3\pi\beta_{p-n}}{2}\frac{1}{L^3}\sum_{\overrightarrow{q}\neq0}Q\left(\frac{(\Lambda^v_M)^2}{(\Lambda^v_M)^2 + Q^2}\right)^n\
\end{equation}
with $n=3,\ 4$. Fitting the discrepancy between the curves and the RBC Collaboration data in Fig.~\ref{fig:chiral-extrapolation}
by adjusting $\beta_{p-n}$ leads to the extracted values for the 
isovector magnetic polarizability $\beta_{p-n}$ at each pion mass,
shown in Tab.~\ref{tab:beta-mpi}.

\begin{figure}[t]
\centering\includegraphics[width=10.0cm,clip=true,angle=0]{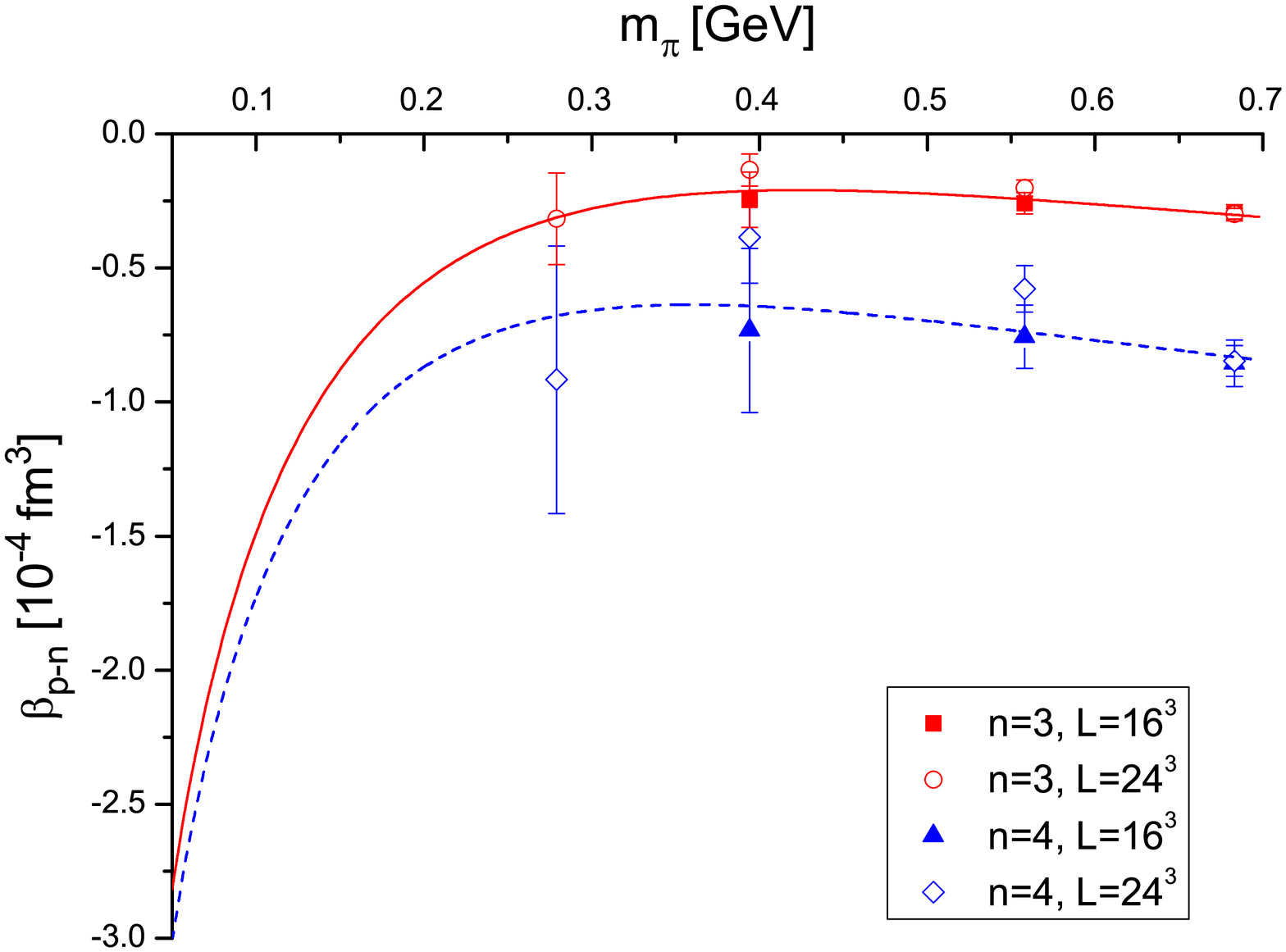}
\caption{Fit results for $\beta_{p-n}$ to the extracted values given by Tab.~\ref{tab:beta-mpi}. Red-solid and blue-dashed curves correspond to the sets of $n=3$ and $n=4$, respectively.}
\label{fig:betapnfit}
\end{figure}
%
\begin{table*}[t]
\begin{center}
\caption{\label{tab:betafit-parameter}Fitted parameters and extrapolated $\beta_{p-n}$ at physical pion mass, in units of $10^{-4}\ \mathrm{fm}^3$. }
\renewcommand{\arraystretch}{1.3}
\begin{tabular}{c|ccc|c}\hline
  \       &       $c_0$        &       $c_1$         &   $\chi^2_{d.o.f}$   & $\beta^{phy}_{p-n}$ \\ \hline
$n=3$     &  $4.83 \pm 0.12$   &   $-6.88 \pm 0.27$  &   $8.19/(7-2)=1.64$  & $-0.98 \pm 0.12$ \\
$n=4$     &  $4.68 \pm 0.34$   &   $-7.69 \pm 0.78$  &   $8.68/(7-2)=1.74$  & $-1.25 \pm 0.36$ \\ \hline
\end{tabular}
\end{center}
\end{table*}

The nucleon electromagnetic polarizabilities have been investigated 
in heavy baryon chiral perturbation theory~\cite{Meissner:1991,Meissner:1993}.
The quantity $\beta_{p-n}$ does not depend on the unknown low energy 
constants $c_2$ and $c^{+}$, which appear in the separate expressions for 
$\beta_p$ and $\beta_n$.
The $1/m_{\pi}$ terms in $\beta_p$ and $\beta_n$ cancel each other. 
Thus we finally obtain: 
\begin{equation}
\beta_{p-n}(m_{\pi})=c_l \ln\frac{m_{\pi}}{M_N} + c_0 + c_1 \frac{m_{\pi}}{M_N}\ ,
\end{equation}
with the model independent coefficient, $c_l$, fixed by chiral perturbation 
theory~\cite{Meissner:1993},
\begin{equation}
c_l = \frac{\alpha g_A^2}{4\pi^2 m_N f_{\pi}^2}(1 + \kappa_s) = 2.51\times 10^{-4}\ \mathrm{fm}^3\ ,
\end{equation}
The other two parameters, determined by fitting the results given 
in Tab.~\ref{tab:beta-mpi}, are summarised in Tab.~\ref{tab:betafit-parameter} 
and the results of those fits are illustrated in Fig.~\ref{fig:betapnfit}.

The physical values for $\beta_{p-n}$, obtained by extrapolating to 
the physical pion mass,
are also shown in the last column of Tab.~\ref{tab:betafit-parameter}. 
It is remarkable that even though the values of $\beta_{p-n}$ found at 
each value of the pion mass tend to be systematically smaller for 
the cubic form factor than for the quartic form factor, 
the values deduced at the physical pion mass are in fairly good 
agreement within their respective uncertainties.
We make a conservative estimate by taking the average value of these two results,
\begin{equation}
\beta_{p-n} = (-1.12 \pm 0.40)\times 10^{-4}\ \mathrm{fm}^3\ .
\end{equation}
This value is of the right sign and order of magnitude compared with 
the experimental result~\cite{McGovern:2012}, 
albeit with a significant smaller error.

In the infinite volume limit, the inelastic subtraction term contributes 
to the electromagnetic p-n mass splitting as
\begin{subequations}
\begin{align}
\label{eq:inelsub3}
\delta M^{inel}_{sub}|_{p-n} &= -\frac{3\beta_{p-n}}{8\pi}\int_0^{\infty}dQ^2 Q^2 \left( \frac{(\Lambda^v_M)^2}{(\Lambda^v_M)^2 + Q^2} \right)^3\no\\
&= 0.30\pm0.04\ \mathrm{MeV}\ ,\\
\label{eq:inelsub4}
\delta M^{inel}_{sub}|_{p-n} &= -\frac{3\beta_{p-n}}{8\pi}\int_0^{\infty}dQ^2 Q^2 \left( \frac{(\Lambda^v_M)^2}{(\Lambda^v_M)^2 + Q^2} \right)^4\no\\
&= 0.12\pm0.04\ \mathrm{MeV}\ .
\end{align}
\end{subequations}
Again, we take the conservative approach of averaging these two results:  
\begin{equation}\label{eq:dMinel}
\delta M^{inel}_{sub}|_{p-n} = 0.21 \pm 0.11\ \mathrm{MeV}\ ,
\end{equation}
where the dominant source of uncertainty comes from the model dependence 
arising from the choice of a cubic or quartic form factor 
in Eq.~(\ref{eq:inelastic-subtraction-term}).
Combining this with Eq.~(\ref{eq:3parts}), we finally obtain the total 
electromagnetic contribution to the proton-neutron mass splitting,
\begin{equation}\label{eq:EMresult}
\delta M^{\gamma}_{p-n} = 1.04 \pm 0.11\ \mathrm{MeV}\ .
\end{equation}

In summary, we have carried out an analysis of the RBC lattice simulations 
of the electromagnetic proton-neutron mass splitting using the 
modified Cottingham sum rule of WLMC. This provides an improved constraint 
on the inelastic subtraction term, which was the major source of uncertainty 
in their work. The isovector nucleon magnetic polarizability was extracted 
as a function of pion mass. The physical value, obtained by chiral 
extrapolation to the physical pion mass, showed a significant improvement 
in precision in comparison with the current experimental value. 
Consequently, we were able to obtain the more accurate result for the 
overall electromagnetic contribution to the proton-neutron 
mass difference given in Eq.~(\ref{eq:EMresult})~\cite{Borsanyi:2014jba}.
This in turn allows us to deduce a more accurate value for the size of the 
contribution to the proton-neutron mass difference arising from the 
difference of the masses of the up and down quarks, namely 
$\delta M^{d-u} \, = \, 2.33 \pm 0.11$ MeV. 
It will be fascinating to 
explore the consequences of this new constraint.

\section*{Acknowledgements}
This work was supported by the University of Adelaide and by the 
Australian Research Council through the ARC Center of Excellence for Particle 
Physics at the Terascale and through grants FL0992247 (AWT) 
and DP110101265 and FT120100821 (RDY).

\end{document}